\documentclass{article}

\usepackage{arxiv}
\usepackage[utf8]{inputenc}
\usepackage[T1]{fontenc}
\IfFileExists{newtxtext.sty}{%
  \usepackage{newtxtext}\usepackage[varg]{newtxmath}}{%
  \usepackage{mathptmx}}
\usepackage{amsmath}
\usepackage{graphicx}
\usepackage{booktabs}
\usepackage{multirow}
\usepackage{array}
\usepackage{tabularx}
\usepackage{microtype}
\usepackage{xcolor}
\usepackage{tikz}
\usetikzlibrary{arrows.meta,decorations.pathreplacing,positioning}
\usepackage[numbers,sort&compress]{natbib}
\usepackage{xurl}
\usepackage[colorlinks=true,linkcolor=blue!50!black,citecolor=blue!50!black,
  urlcolor=blue!50!black]{hyperref}
\usepackage[capitalise,noabbrev]{cleveref}

\newcommand{\cfg}[1]{\textsf{#1}}
\newcommand{\hd}[1]{\begin{tabular}[b]{@{}c@{}}#1\end{tabular}}

\title{Accelerating QUIC Encryption in WebAssembly Using WASM-SIMD}
\date{}

\author{
  Sota Yamawaki \\
  Graduate School of Informatics\\
  Kyoto University\\
  Yoshida-honmachi, Sakyo-ku, Kyoto 606-8501, Japan \\
  \texttt{yamawaki.souta.36p@st.kyoto-u.ac.jp} \\
  \And
  Yasuo Okabe \\
  Academic Center for Computing and Media Studies\\
  Kyoto University\\
  Yoshida-honmachi, Sakyo-ku, Kyoto 606-8501, Japan \\
  \texttt{okabe@i.kyoto-u.ac.jp} \\
}

\renewcommand{\shorttitle}{Accelerating QUIC Encryption in WebAssembly Using WASM-SIMD}

\hypersetup{
  pdftitle={Accelerating QUIC Encryption in WebAssembly Using WASM-SIMD},
  pdfsubject={cs.NI, cs.PF},
  pdfauthor={Sota Yamawaki, Yasuo Okabe},
  pdfkeywords={WebAssembly, QUIC, WASM-SIMD, authenticated encryption},
}

\begin{document}
\maketitle

{\renewcommand{\thefootnote}{}%
\footnotetext{A Japanese version of this work has been submitted as a
  manuscript for presentation at the IEICE Technical Committee on Internet
  Architecture (IA).
  The abstract is available at
  \url{https://ken.ieice.org/ken/paper/202611028cZo/eng/}.}}

\begin{abstract}
WebAssembly (Wasm) runs untrusted code in a portable, lightweight sandbox and
is increasingly used to host services on shared servers and edge platforms.
Running the QUIC transport protocol inside Wasm would let each application
choose and update its own transport without changing the host. QUIC, however,
encrypts every packet, and Wasm offers neither dedicated AES instructions nor
carry-less multiplication; a Wasm QUIC stack has been reported to reach only
17.6\% of native goodput, with cryptography as its largest cost. Whether the
128-bit SIMD extension of Wasm (WASM-SIMD) can recover this cost, and what
limits a pure-Wasm stack once it does, has remained unclear. Here we show that
explicit WASM-SIMD kernels recover much of the cryptographic cost, after which
non-cryptographic costs become the limit. On two hosts connected by 10-Gigabit
Ethernet, kernels that process eight AES blocks or four ChaCha20 blocks per
pass raise receiver goodput over a non-SIMD baseline by 61.34\% for
AES-128-GCM and 31.55\% for ChaCha20-Poly1305, whereas compiler
auto-vectorization gains only 1.28\% for AES-128-GCM. ChaCha20-Poly1305 still
delivers 1.99 times the goodput of AES-128-GCM, confirming end to end the
preference that stand-alone cipher benchmarks suggest. Receiver-side profiling
combined with Amdahl's law predicts that even native-speed cryptography would
leave Wasm at roughly one third of the goodput of the fastest native stack.
Cryptographic acceleration is therefore necessary but not sufficient: memory
allocation, runtime overhead and per-datagram UDP I/O now bound pure-Wasm
transports. We anticipate that this per-category cost breakdown will guide
further optimization of Wasm network stacks and the design of system
interfaces for batched I/O.
\end{abstract}


\section{Introduction}

Running the entire QUIC transport~\cite{rfc9000,langley2017} inside
WebAssembly (Wasm) lets each application on a shared server or edge platform
replace and update its transport without modifying the
host~\cite{water2024}. Wasm and the WebAssembly System Interface (WASI) offer
such platforms language independence and lightweight
sandboxing~\cite{gackstatter2022}, and recent work composes QUIC stacks from
isolated Wasm components~\cite{spies2025}.

This flexibility costs performance, mainly in cryptography, because Wasm lacks
the instructions that make encryption cheap on native platforms. QUIC applies
authenticated encryption to every packet. On native platforms,
dedicated AES instructions keep this cost small: disabling packet protection
improves performance by only 10--20\%~\cite{kempf2024}, and the throughput gap
between QUIC and TCP on fast links stems mainly from packet processing rather
than encryption~\cite{zhang2024}. Wasm, in contrast, prioritizes portability
and lacks both dedicated AES and carry-less multiplication
instructions~\cite{wasmsimd}. Spies et al.~\cite{spies2025} found that their
fastest Wasm configuration reached only 17.6\% of the goodput
of the fastest native configuration, with
cryptography as the largest contributor to the gap.

Without AES instructions, the usual preference for AES-GCM also becomes
questionable. TLS~1.3 requires AES-128-GCM and
recommends AES-256-GCM and ChaCha20-Poly1305 among its authenticated
encryption with associated data (AEAD) algorithms~\cite{rfc8446}, and QUIC uses
the same cipher suites~\cite{rfc9001}. Which of AES-128-GCM and
ChaCha20-Poly1305 is faster depends on whether the platform provides AES
instructions~\cite{jaeger2023}, so the native ranking cannot be assumed in Wasm.

Delegating cryptography to the host would sidestep the problem but undermine
the reason for a Wasm transport. A host interface such as
WASI-Crypto~\cite{wasicrypto}, or TLS termination by the host, ties the
transport to features that every platform must provide, moves key material
across the sandbox boundary, and, for QUIC, requires a guest--host call per
packet. A pure-Wasm stack avoids these dependencies but must make cryptography
fast with the instructions that Wasm offers.

Existing work leaves two questions open: whether WASM-SIMD, the 128-bit
single-instruction-multiple-data extension of Wasm, can recover a substantial
part of the cryptographic cost, and what limits a pure-Wasm QUIC stack once it
does. The first is unsettled because the Wasm build of BoringSSL used by Spies
et al.\ was a generic implementation with assembly disabled, and stand-alone
benchmarks of Wasm ciphers~\cite{denis2026} do not show how cipher speed
translates into the goodput and CPU time of a complete QUIC transfer. The
second requires attributing CPU time to processing categories, not only
measuring throughput.

We answer both questions with explicit WASM-SIMD kernels in a pure-Wasm QUIC
stack that targets WASI~0.3, evaluated on two hosts connected by 10-Gigabit
Ethernet (10GbE). The underlying techniques, bitsliced~\cite{kasper2009} and
fixsliced~\cite{fixslicing} AES and vectorized ChaCha20~\cite{gollgueron2014},
are known; we adapt them to WASM-SIMD and quantify their end-to-end effect on
QUIC. Our contributions are threefold:

\begin{enumerate}
\item \textbf{Cryptographic kernels tailored to WASM-SIMD.} An eight-block
  fixsliced AES kernel and a four-block ChaCha20 kernel with shuffle-based
  rotations raise goodput over the non-SIMD baseline by 61.34\% with
  AES-128-GCM and 31.55\% with ChaCha20-Poly1305. Auto-vectorization alone
  gains only 1.28\% with AES-128-GCM.
\item \textbf{An end-to-end comparison of ciphers in pure Wasm.}
  ChaCha20-Poly1305 delivers 2.44 times the goodput of AES-128-GCM without
  SIMD and 1.99 times with our kernels. This turns the preference suggested by
  stand-alone benchmarks~\cite{denis2026} into an end-to-end measurement.
\item \textbf{A quantification of the costs that remain.} A per-category
  breakdown of receiver CPU time and Amdahl's law show that even native-level
  cryptography would leave Wasm at about one third of the fastest native
  goodput. Memory allocation, runtime overhead, and per-datagram socket I/O
  emerge as the next targets.
\end{enumerate}

\paragraph{Scope.}
We study the receiving side of a bulk transfer over a single, lossless QUIC
connection between two x86-64 hosts, with Wasmtime as the runtime and
AES-128-GCM and ChaCha20-Poly1305 as the AEADs. We do not evaluate Wasm
senders, AES-256-GCM, short connections, lossy or high-latency paths, or other
runtimes and instruction sets; \cref{sec:discussion} discusses what our
conclusions do and do not cover.

\Cref{sec:background,sec:related} give background and related work.
\Cref{sec:stack,sec:method} describe our stack, the evaluated configurations,
and the methodology. \Cref{sec:profile} profiles the baseline,
\cref{sec:impl} presents the kernels, and \cref{sec:eval} evaluates them.
\Cref{sec:discussion} discusses limitations, and \cref{sec:conclusion}
concludes.

\section{Background}
\label{sec:background}

\subsection{WebAssembly and WASI}

Wasm~\cite{haas2017} is a sandboxed, stack-based instruction set, and
WASI~\cite{wasi} exposes host functionality to it outside the browser. A Wasm
module runs isolated from the host in its own linear memory and interacts with
the host only through explicit imports and exports. WASI defines host
functionality, such as sockets and clocks, as type-safe interfaces via the
Component Model and WebAssembly Interface Types (WIT). WASI~0.3 (formerly
Preview~3)~\cite{wasip3}, which this paper targets, adds asynchronous
functions to the Component Model and makes socket send and receive
asynchronous; the guest waits for outstanding operations by grouping
them in a \emph{waitable set}.

Wasm adds costs that native code does not pay, both at the host boundary and
in computation. Every host-function call passes through the runtime, and a
received packet may be copied from host-side buffers into linear memory in
addition to the copy from the OS kernel to user space. Even compute-bound code
runs slower: with default bounds checking, performance-oriented runtimes
execute PolyBench/C kernels within 20\% of native execution time on x86-64 and
within 35\% on Armv8~\cite{szewczyk2022}. A comparison between Wasm and native
execution must therefore account for memory operations and host calls as well
as computation.

\subsection{QUIC and Authenticated Encryption}

QUIC~\cite{rfc9000} runs over the UDP and protects
every packet with authenticated encryption keyed by the TLS~1.3 key
schedule~\cite{rfc9001}. A separate header protection additionally masks part
of the header, including the packet number. Initial packets are always
protected with AES-128-GCM~\cite{rfc9001}.

The two AEADs we compare use different ciphers but structurally similar MACs.
AES-128-GCM~\cite{nist80038d} combines AES-128 in counter (CTR) mode with
GHASH, a universal hash over $\mathrm{GF}(2^{128})$. Each AES round applies
byte substitution through S-boxes (SubBytes), a byte permutation (ShiftRows), a
matrix multiplication over a finite field (MixColumns), and a key addition.
ChaCha20-Poly1305~\cite{rfc8439} combines the ChaCha20 stream cipher, which uses
only 32-bit additions, rotations, and XORs (ARX), with the Poly1305 message
authentication code (MAC). Both MACs evaluate a polynomial, over
$\mathrm{GF}(2^{128})$ for GHASH and modulo the prime $2^{130}-5$ for
Poly1305.

\subsection{SIMD and WASM-SIMD}

SIMD accelerates computations that apply
the same operation to independent data. It splits a vector register into lanes
and applies one instruction to all of them. Computations that depend on the
immediately preceding result, carry across lanes, branch on data, or access
memory non-contiguously are hard to vectorize.

Cryptographic code can exploit SIMD across blocks or across bits, through
either the compiler or hand-written kernels. Compiler auto-vectorization needs
no code changes, whereas explicit kernels design the data layout and operation
sequence for the vector unit. Kernels either process multiple blocks under the
same key in parallel or distribute a bit-level state across lanes. The latter, \emph{bitslicing}, gathers the $i$-th bits of many blocks
in one word so that the S-box can be evaluated with logic operations instead of
table lookups.

WASM-SIMD~\cite{wasmsimd} is narrower and less complete than native SIMD
extensions, and these gaps shape our kernels in \cref{sec:impl}. Its vector
width is fixed at 128 bits, compared with 256 and 512 bits for AVX2 and
AVX-512. It offers no AES instructions, no carry-less multiplication, and no
lane-wise rotate. Byte permutations are available as \texttt{i8x16.shuffle},
whose mask is a compile-time constant, and \texttt{i8x16.swizzle}, whose mask
is a run-time value.

\section{Related Work}
\label{sec:related}

\paragraph{QUIC and network stacks in Wasm.}
Prior work on Wasm network stacks has identified cryptography and I/O as
bottlenecks but has not accelerated cryptography with WASM-SIMD.
WATER~\cite{water2024} distributes censorship-circumvention transports as Wasm
modules that can be updated without modifying the application; its appendix
notes that cryptography in Wasm is slower than native for lack of hardware
acceleration. Spies et al.~\cite{spies2025} composed a QUIC stack from Wasm
components and evaluated an HTTP/3 server based on quiche and BoringSSL over
the loopback path of a single machine. Their CPU-cycle breakdown identified
cryptography, memory copies, and socket I/O through WASI as the main
bottlenecks, but their Wasm BoringSSL was a generic build with assembly
disabled. Aoyagi and Okabe~\cite{aoyagi2026} batched multiple datagrams into a
single list per UDP send and receive call within standard WASI Preview~2,
raising the goodput of a quiche-based HTTP/3 server on 10GbE from 13.2\% to
15.5\% of native; our work on cryptography is complementary. The
polymorph-tls project~\cite{polymorph} provides a pure-Wasm TLS~1.3
implementation with a QUIC compatibility layer and reports that fixsliced
AES-128-GCM with software GHASH is much slower than ChaCha20-Poly1305 at QUIC
packet sizes. We found neither SIMD kernels nor an end-to-end receiver CPU
profile in that project.

\paragraph{QUIC performance and the cost of cryptography.}
On native platforms, where hardware AES is available, cryptography is a
secondary cost of QUIC. QUIC runs congestion control, retransmission, and other
tasks in user space, so CPU processing can limit throughput on fast
links~\cite{langley2017}. Zhang et al.~\cite{zhang2024} showed that QUIC can be
up to 45.2\% slower than TCP+TLS+HTTP/2 on fast links and attributed this to
receiver-side packet processing rather than encryption. K\"onig et
al.~\cite{koenig2025} found that QUIC implementations still fall short of TCP
on 10GbE, often because of per-packet processing. Kempf et al.~\cite{kempf2024}
measured a 10--20\% gain from disabling packet protection and found AES-based
header protection inexpensive. Jaeger et al.~\cite{jaeger2023} showed that
packet I/O becomes the main load in native QUIC with AES-NI or ChaCha20. In
the versions of LSQUIC and quiche they evaluated (January 2023), enabling
Generic Segmentation Offload (GSO) and its receive counterpart did not improve
goodput, and application-side support for batched I/O was limited; enabling
offloads alone thus did not remove the packet I/O bottleneck. Mukherjee et al.~\cite{mukherjee2025} observed that the choice of
cryptographic library substantially affects QUIC throughput, and
Oku~\cite{oku2020} showed that libraries not optimized for packet-sized inputs
inflate the AEAD cost of QUIC. Without hardware AES, Wasm shifts this balance,
and this is the regime we study.

\paragraph{Cryptography in Wasm.}
Prior work on cryptography in Wasm has addressed verification, constant-time
guarantees, host delegation, and stand-alone speed, but not end-to-end
transport performance. Protzenko et al.~\cite{protzenko2019} compiled the
verified HACL$^\ast$ library to Wasm before WASM-SIMD was standardized.
CT-Wasm~\cite{watt2019} and the constant-time WebAssembly
proposal~\cite{constanttime} add secret types and instructions for
constant-time cryptography. WASI-Crypto~\cite{wasicrypto} lets Wasm code call
the host's native libraries and accelerators; our design instead keeps
cryptography inside Wasm and needs no special API. Denis's
benchmarks~\cite{denis2026} show that pure-Wasm AES-128-GCM reaches 351~Mb/s on
the Fastly platform, slower than the AEGIS and HiAE ciphers on several content
delivery platforms, and that on Wasmtime hand-optimized ChaCha20-Poly1305
reaches 492~MiB/s, whereas AES-128-GCM reaches 46--90~MiB/s. We measure how
such stand-alone differences translate into the CPU time and goodput of QUIC.

\paragraph{SIMD acceleration of cryptography.}
Our kernels adapt established SIMD techniques to the restrictions of WASM-SIMD.
Bitsliced AES-GCM~\cite{kasper2009} and vectorized
ChaCha20~\cite{gollgueron2014} parallelize each cipher along its computational
structure, and HACL$\times$N~\cite{polubelova2020} generalizes the
vectorization of ChaCha20 and Poly1305 to 128-, 256-, and 512-bit vectors.
K\"asper and Schwabe's bitsliced AES already processes eight blocks in 128-bit
registers on x86-64. Fixslicing~\cite{fixslicing} arranges the state so that
repeated ShiftRows computations can be omitted, reducing the operation count of
the linear layer by 52\% on 32-bit platforms compared with the previously
fastest bitsliced implementation. We combine the eight-block structure of the
former with the fixsliced linear layer of the latter, which avoids repeated
byte permutations.

\section{A Pure-Wasm QUIC Stack}
\label{sec:stack}

Our \emph{pure-Wasm QUIC stack} runs QUIC, TLS, and cryptography in a single
Wasm module without native libraries. We base it on Quinn~\cite{quinn}, whose
standard configuration relies on the tokio asynchronous runtime and on the
assembly-optimized ring library~\cite{ring} and therefore does not compile for
WASI~0.3 without modification. We instead use quinn-proto, Quinn's sans-I/O
QUIC state machine, and drive it with our own event loop over the asynchronous
\texttt{send} and \texttt{receive} functions of \texttt{wasi:sockets} and the
timers of \texttt{wasi:clocks}. All QUIC processing runs inside the
asynchronous export of \texttt{wasi:cli/run}, and the build target is
\texttt{wasm32-wasip3}.

The event loop keeps one receive outstanding at all times and yields to the
host only when no datagram awaits processing. It then waits until a receive, a
send, or a timer completes. Each send or receive call handles a single
datagram.

For cryptography, we use the pure-Rust RustCrypto~\cite{rustcrypto} through
rustls-rustcrypto~\cite{rustlsrc} rather than a Wasm port of ring. Ports such
as ring-wasi~\cite{wasixcrypto} make ring available in extended environments
such as WASIX, but its native assembly optimizations do not carry over to Wasm.
RustCrypto contains no assembly and keeps all cryptography within standard
WASI~0.3; we optimize it with explicit WASM-SIMD kernels. We also added QUIC
header protection and Initial-packet encryption, which the provider did not
previously support.

\subsection{Evaluated Configurations}
\label{sec:configs}

\begin{figure}[t]
  \centering
  \includegraphics[width=0.78\linewidth]{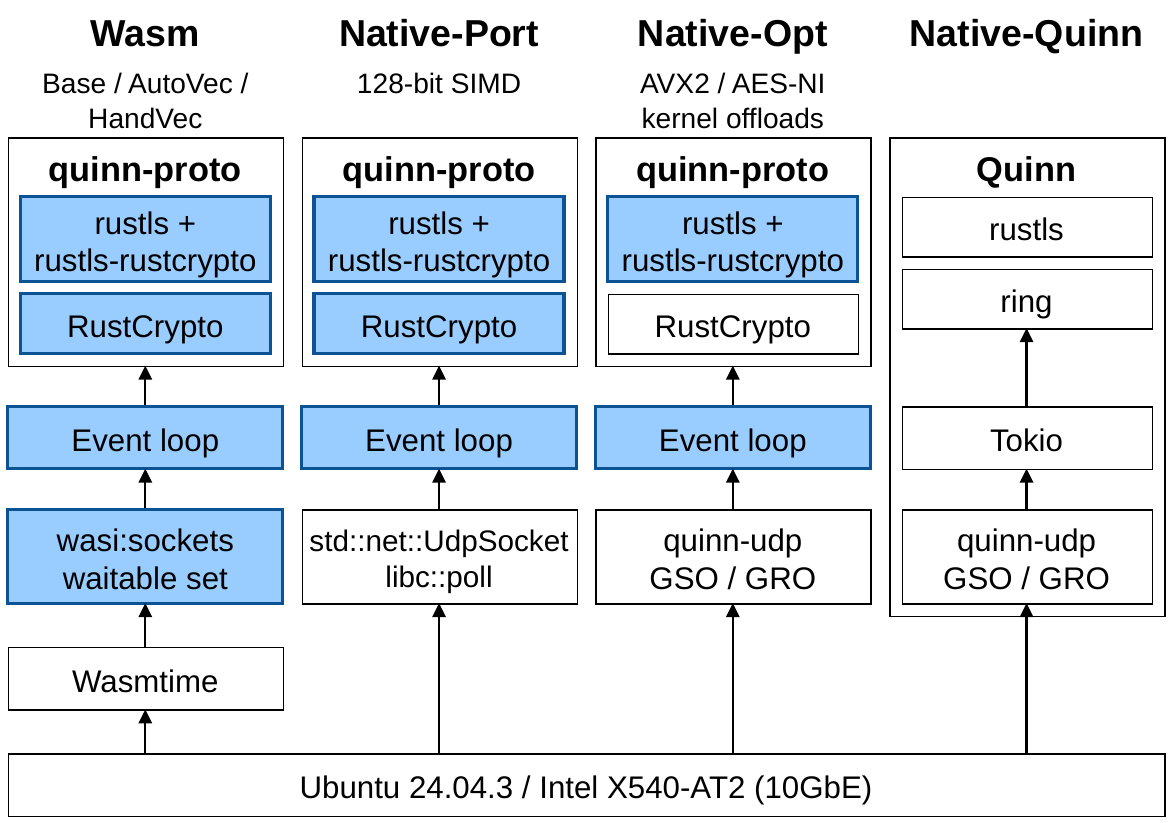}
  \caption{Evaluated stacks and the receive path. The Wasm column combines
    three configurations. Blue blocks indicate components we implemented or
    extended.}
  \label{fig:stack}
\end{figure}

We compare six builds, each with two AEADs, ordered so that each successive
pair isolates one source of the gap between Wasm and native
(\cref{fig:stack}). In every configuration, Initial packets use AES-128-GCM.

\begin{description}
\item[\cfg{Wasm-Base}] does not use WASM-SIMD and serves as the baseline.
\item[\cfg{Wasm-AutoVec}] enables WASM-SIMD and relies solely on compiler
  auto-vectorization.
\item[\cfg{Wasm-HandVec}] adds the explicit kernels of \cref{sec:impl}. The
  MACs remain scalar (\cref{sec:mac}).
\item[\cfg{Native-Port}] is the \cfg{Wasm-HandVec} code built natively with the
  native socket API. It keeps 128-bit SIMD and uses no kernel offloads, so the
  step from \cfg{Wasm-HandVec} isolates the cost of the Wasm execution
  environment.
\item[\cfg{Native-Opt}] extends \cfg{Native-Port} with AVX2 (256-bit SIMD),
  AES-NI (dedicated AES instructions), and the kernel offloads GSO and Generic
  Receive Offload (GRO), which process multiple UDP datagrams per system call.
  The step from \cfg{Native-Port} isolates host features that Wasm cannot use.
\item[\cfg{Native-Quinn}] is the official Quinn on tokio and ring, the fastest
  native configuration in our evaluation. The step from \cfg{Native-Opt}
  isolates differences in implementation maturity.
\end{description}

\section{Experimental Methodology}
\label{sec:method}

We measure receiver-side goodput and CPU time for 1-GiB bulk transfers between
two hosts connected directly by 10GbE (\cref{tbl:testbed}). The server sends
1~GiB of application data to the client. Let $B$ be the number of bytes
delivered to the receiving application without duplication and $t$ the
transfer time in seconds; goodput is $G = 8B / (10^{6} t)$~Mb/s. CPU time is
the \texttt{task-clock} of all threads of the receiving process, measured with
Linux \texttt{perf stat} and normalized to ms/GiB.

\begin{table}[t]
  \caption{Testbed. Goodput and CPU time are each measured over 20 runs, and CPU
    profiles are sampled at 4,000~Hz over 5 runs. The receiving process is
    pinned to CPUs 6--9. Wasmtime runs with
    \texttt{-S inherit-network -S udp -S p3 -W component-model-async}, plus
    \texttt{-{}-profile=perfmap} when profiling. MTU: maximum transmission unit.}
  \label{tbl:testbed}
  \centering
  \small
  \newcolumntype{Y}{>{\raggedright\arraybackslash}X}
  \newcommand{\both}[1]{\multicolumn{2}{>{\hsize=\dimexpr2\hsize+2\tabcolsep\relax\raggedright\arraybackslash}X@{}}{#1}}
  \begin{tabularx}{0.85\linewidth}{@{}lYY@{}}
    \toprule
     & Receiver & Sender \\
    \midrule
    CPU & Ryzen 9 5900X & Core i9-12900K \\
    Memory & 32 GiB & 16 GiB \\
    Network card & Intel X540-AT2 & Intel X550 \\
    OS & Ubuntu 24.04.3 & Ubuntu 26.04.1 \\
    \midrule
    Link & \both{Direct 10GbE connection} \\
    OS kernel & \both{Linux 7.0.0-31-generic, \texttt{performance} CPU governor} \\
    Buffers & \both{Socket 8/32 MiB (default/max), backlog 30,000} \\
    Software & \both{Wasmtime 49.0.2, rustc 1.100.0-nightly, perf 7.0.14} \\
    QUIC & \both{quinn-proto 0.11.16, rustls 0.23.42, rustls-rustcrypto 0.0.2-alpha} \\
    Protocol & \both{CUBIC congestion control~\cite{rfc9438}, MTU 1,500 bytes} \\
    \bottomrule
  \end{tabularx}
\end{table}

\paragraph{Receiver focus.}
We measure the receiver because receiver-side processing limits the data rate
of QUIC on fast links~\cite{zhang2024}. ChaCha20 and the CTR part of
AES-128-GCM generate the same keystream for encryption and decryption, so we
expect our kernels to cost about the same on the sender; we do not evaluate
Wasm senders.

\paragraph{Socket buffers.}
We enlarged the socket buffers (\cref{tbl:testbed}) so that packet drops would
not mask CPU limits. With the default UDP receive buffer, overflow dropped
packets at the receiving socket, and QUIC congestion control then lowered the
sending rate before the receiver reached its CPU limit. With the enlarged
buffers, preliminary measurements showed no overflow.

\paragraph{Sender.}
Every experiment uses \cfg{Native-Quinn} as the sender so that the sender
never limits receiver goodput. Pairing each receiver with a sender of the same
configuration would couple a slow receiver with a similarly slow sender and
confound their limits. The sender's CPU utilization is 6--26\% in the Wasm
measurements and at most 52\% with \cfg{Native-Port} and \cfg{Native-Opt}. It
approaches saturation (85--89\%) only with \cfg{Native-Quinn} and
ChaCha20-Poly1305.

\paragraph{Is the receiver CPU-bound?}
The Wasm receivers are CPU-bound, so their goodput is inversely proportional
to CPU time; the native receivers are not. \Cref{tbl:cpubound} compares the
time needed to transfer 1~GiB at the measured goodput with the receiver CPU
time per GiB. For all three Wasm builds and both algorithms, CPU time exceeds
transfer time by 0.35--0.46~s, so the guest thread is busy throughout the
transfer. The nearly constant excess suggests a fixed per-process cost of
roughly 0.4~s, such as module compilation, that counts toward CPU time but not
transfer time. For \cfg{Native-Port} and \cfg{Native-Opt}, CPU time is only
68--92\% of transfer time, so something other than receiver CPU limits their
goodput (\cref{sec:limit-theory,sec:discussion}).

\begin{table}[t]
  \caption{Transfer time for 1~GiB at the measured goodput ($8\,589.9\ \mathrm{Mb} / G$)
    compared with the receiver CPU time per GiB. Goodputs other than those of
    \cfg{Wasm-Base} and \cfg{Wasm-HandVec} are the rounded values shown in
    \cref{fig:main}; the last column is derived from these numbers.}
  \label{tbl:cpubound}
  \centering
  \small
  \begin{tabular}{@{}llrrrr@{}}
    \toprule
    Algorithm & Build & Goodput (Mb/s) & Transfer time (s) & CPU time (s) & CPU/transfer \\
    \midrule
 \multirow{6}{*}{\hd{AES-128-\\GCM}} & \cfg{Wasm-Base} & 612.4 & 14.03 & 14.48 & 1.03 \\
  & \cfg{Wasm-AutoVec} & 620 & 13.85 & 14.31 & 1.03 \\
  & \cfg{Wasm-HandVec} & 988.1 & 8.69 & 9.09 & 1.05 \\
  & \cfg{Native-Port} & 1,235 & 6.96 & 6.42 & 0.92 \\
  & \cfg{Native-Opt} & 5,131 & 1.67 & 1.14 & 0.68 \\
  & \cfg{Native-Quinn} & 9,081 & 0.95 & 0.90 & 0.95 \\
    \midrule
 \multirow{6}{*}{\hd{ChaCha20-\\Poly1305}} & \cfg{Wasm-Base} & 1,493.8 & 5.75 & 6.14 & 1.07 \\
  & \cfg{Wasm-AutoVec} & 1,727 & 4.97 & 5.36 & 1.08 \\
  & \cfg{Wasm-HandVec} & 1,965.1 & 4.37 & 4.73 & 1.08 \\
  & \cfg{Native-Port} & 2,514 & 3.42 & 2.84 & 0.83 \\
  & \cfg{Native-Opt} & 3,994 & 2.15 & 1.60 & 0.74 \\
  & \cfg{Native-Quinn} & 7,345 & 1.17 & 1.11 & 0.95 \\
    \bottomrule
  \end{tabular}
\end{table}

\paragraph{Profile classification.}
We attribute CPU time to processing categories by matching the symbol name of
each profile sample, and where necessary its dynamic shared object (DSO) name,
against regular expressions. Just-in-time (JIT) compiled code from
Wasmtime~\cite{wasmtime} is mapped to symbols via \texttt{perfmap}, kernel-space
symbols are resolved as well, and shares are weighted by the sampling period.
Of our nine categories, \cref{tbl:cycles} omits \texttt{ack-loss}
(acknowledgment, loss detection, and congestion control), which stays below 4~ms/GiB in every
configuration, and merges QUIC state updates and frame/header encoding and
decoding into \texttt{quic-proto}; totals include all nine. Because categories
are inferred from names, some samples may be misclassified. Our
\texttt{crypto}, \texttt{copy}, and \texttt{socket-io} map onto the
cryptography, memory copy, and socket I/O categories of Spies et
al.~\cite{spies2025}, and the sum of the remaining categories maps onto their
other.

\section{Profiling the Baseline}
\label{sec:profile}

Cryptography dominates the receiver CPU time of the non-SIMD baseline,
accounting for 81.3\% with AES-128-GCM and 55.4\% with ChaCha20-Poly1305
(\cref{tbl:baseline}). These shares make cryptography the first target for
acceleration.

\begin{table}[t]
  \caption{Baseline (\cfg{Wasm-Base}) receiver goodput and CPU time per GiB. The
    non-crypto column is the total minus the crypto column.}
  \label{tbl:baseline}
  \centering
  \small
  \begin{tabular}{@{}lrrrrr@{}}
    \toprule
    Algorithm & Goodput (Mb/s) & Total (ms/GiB) & crypto & non-crypto & crypto share \\
    \midrule
    AES-128-GCM & 612.4 & 14,483 & 11,768 & 2,715 & 81.3\% \\
    ChaCha20-Poly1305 & 1,493.8 & 6,142 & 3,403 & 2,739 & 55.4\% \\
    \bottomrule
  \end{tabular}
\end{table}

The difference between the two algorithms lies almost entirely in
cryptography. Non-cryptographic CPU time is nearly identical (2,715 and
2,739~ms/GiB), as expected because the number of packets and QUIC processing
steps per GiB does not depend on the algorithm. The cryptographic CPU time of
AES-128-GCM is 3.46 times that of ChaCha20-Poly1305, and correspondingly
ChaCha20-Poly1305 reaches 1,493.8~Mb/s, 2.44 times the 612.4~Mb/s of
AES-128-GCM. AES-128-GCM therefore has the most to gain.

\section{Accelerating Authenticated Encryption with WASM-SIMD}
\label{sec:impl}

We adapt known parallelization
techniques~\cite{gollgueron2014,kasper2009,fixslicing} to the instruction
constraints of WASM-SIMD. For AES, we widen a bit-level state layout to 128-bit
vectors; for ChaCha20, we combine block-level parallelism with byte shuffles
that compensate for the missing rotate instruction.

\subsection{ChaCha20 Kernel}

\begin{figure}[t]
  \centering
\begin{tikzpicture}[
  x=1cm, y=1cm, >={Latex[length=2mm]},
  font=\small,
  cell/.style={draw, minimum height=0.55cm, inner sep=0pt, outer sep=0pt},
  title/.style={font=\bfseries, anchor=west},
  stage/.style={draw, thick, align=center, minimum height=1.05cm, inner sep=3pt},
]
\node[title] at (-0.2, 0.55) {(a) ChaCha20};

\foreach \b in {0,...,3} {
  \pgfmathsetmacro{\y}{-0.4 - 0.7*\b}
  \node[anchor=east] at (0.55, \y) {$B_{\b}$};
  \node[cell, minimum width=0.85cm] at (1.0,  \y) {$x_0$};
  \node[cell, minimum width=0.85cm] at (1.85, \y) {$x_1$};
  \node[cell, minimum width=0.85cm] at (2.7,  \y) {$\cdots$};
  \node[cell, minimum width=0.85cm] at (3.55, \y) {$x_{15}$};
}

\draw[->] (4.2, -1.45) -- (6.0, -1.45)
  node[midway, above, align=center, font=\footnotesize] {interleave\\into lanes};

\foreach \b in {0,...,3}
  \node at (7.35 + 1.5*\b, 0.55) {$B_{\b}$};
\foreach \r/\y in {0/-0.4, 1/-1.1, 15/-2.5} {
  \node[anchor=east] at (6.5, \y) {$R_{\r}$};
  \foreach \b in {0,...,3}
    \node[cell, minimum width=1.5cm] at (7.35 + 1.5*\b, \y) {$x_{\r}$};
}
\node at (9.6, -1.8) {$\vdots$};

\node at (2.0, -3.45) {State matrix (64 bytes)};
\node at (2.0, -4.75) {$\left(\begin{array}{@{\,}cccc@{\,}}
  \mathrm{const} & \mathrm{const} & \mathrm{const} & \mathrm{const}\\
  \mathrm{key}_0 & \mathrm{key}_1 & \mathrm{key}_2 & \mathrm{key}_3\\
  \mathrm{key}_4 & \mathrm{key}_5 & \mathrm{key}_6 & \mathrm{key}_7\\
  \mathrm{ctr}   & n_0 & n_1 & n_2
\end{array}\right)$};

\node[anchor=east] at (6.5, -3.6) {Counter};
\foreach \b/\c in {0/c, 1/c+1, 2/c+2, 3/c+3} {
  \node[cell, minimum width=1.5cm] at (7.35 + 1.5*\b, -3.6) {$\c$};
  \node[font=\footnotesize] at (7.35 + 1.5*\b, -4.2) {\texttt{u32}};
}
\draw[decorate, decoration={brace, mirror, amplitude=5pt}]
  (6.65, -4.5) -- (12.65, -4.5)
  node[midway, below=6pt, align=center]
  {one \texttt{v128} ($4 \times$ \texttt{u32}): one lane per state matrix};

\draw[gray] (-0.3, -6.0) -- (12.8, -6.0);

\node[title] at (-0.2, -6.55) {(b) AES-128 with fixslicing};
\node[font=\bfseries] at (9.65, -6.55) {$8 \times$ \texttt{v128}};

\foreach \b/\y in {0/-7.3, 1/-8.0, 7/-9.4}
  \node[cell, minimum width=3.4cm] at (2.25, \y) {$B_{\b}$};
\node at (2.25, -8.7) {$\vdots$};

\draw[->] (4.2, -8.35) -- (6.0, -8.35)
  node[midway, above, font=\footnotesize] {transpose};

\foreach \r/\y in {0/-7.3, 1/-8.0, 7/-9.4} {
  \node[anchor=east] at (6.5, \y) {$R_{\r}$};
  \foreach \b in {0,...,7}
    \node[cell, minimum width=0.75cm, font=\footnotesize]
      at (6.975 + 0.75*\b, \y) {$B_{\b,\r}$};
}
\node at (9.6, -8.7) {$\vdots$};
\node[anchor=west] at (6.6, -10.05) {$B_{i,j}$: bit plane $j$ of block $i$};

\node[stage, anchor=west] (sbox) at (-0.2, -11.85)
  {256-entry\\table lookup\\$y = S[x]$};
\node[above=1pt of sbox] {AES S-box};
\node[right=1.9cm of sbox] (xin) {$x_0 \ldots x_7$};
\draw[->] (sbox) -- (xin)
  node[midway, above, font=\footnotesize] {replaced by};
\node[stage, minimum width=2.3cm, right=0.5cm of xin] (top)
  {Top\\linear layer\\XOR};
\node[stage, minimum width=2.3cm, right=0.5cm of top] (mid)
  {Nonlinear\\layer\\AND/XOR};
\node[stage, minimum width=2.3cm, right=0.5cm of mid] (bot)
  {Bottom\\linear layer\\XOR/XNOR};
\node[right=0.5cm of bot] (yout) {$y_0 \ldots y_7$};
\draw[->] (xin) -- (top);
\draw[->] (top) -- (mid);
\draw[->] (mid) -- (bot);
\draw[->] (bot) -- (yout);
\draw[decorate, decoration={brace, amplitude=5pt}]
  ([yshift=4pt]top.north west) -- ([yshift=4pt]bot.north east)
  node[midway, above=6pt] {Boyar--Peralta circuit (XOR/AND)};
\end{tikzpicture}
  \caption{Data layouts of the SIMD cryptographic kernels. (a) ChaCha20: the
    words of four blocks are interleaved across lanes. (b) AES-128: eight
    blocks are transposed into bit planes.}
  \label{fig:kernels}
\end{figure}
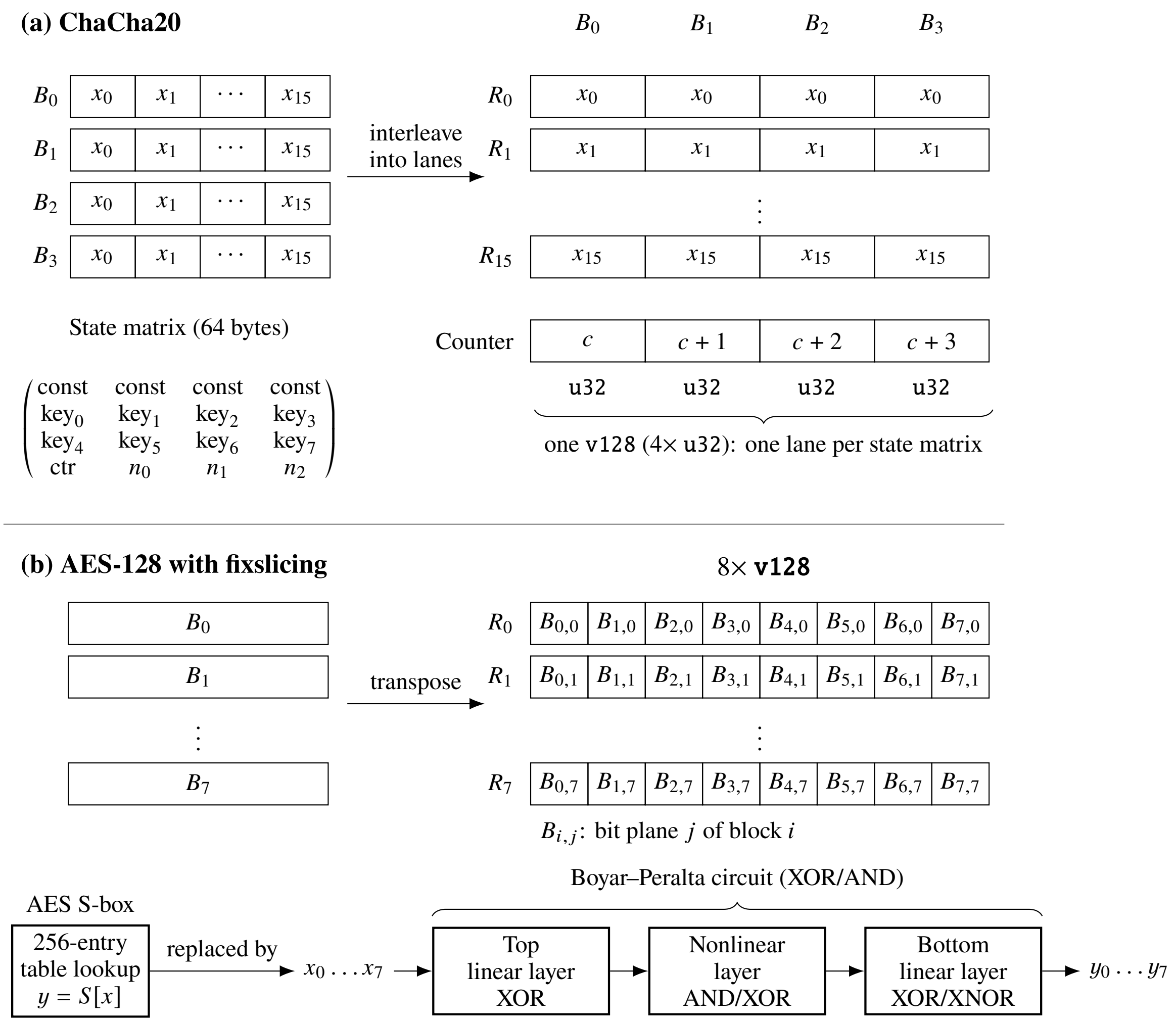

Our ChaCha20 kernel processes four blocks at once, exploiting the independence
of ChaCha20 blocks~\cite{gollgueron2014}. ChaCha20 repeatedly applies ARX
operations to a $4 \times 4$ state of sixteen 32-bit words. On the main path,
each 128-bit register holds the word at the same position in four blocks
(\cref{fig:kernels}(a)). Only the block counter differs between lanes,
initialized as \texttt{u32x4}$(c, c+1, c+2, c+3)$. Sixteen \texttt{v128}
registers thus hold the entire state of four blocks, and the ten double rounds
are fully unrolled without branches. When fewer than four blocks remain, a
row-vector layout places the four words of each state row in one
\texttt{v128}.

Byte shuffles replace two of the four rotations, cutting each from three
instructions to one. Without a lane-wise vector rotate, a rotation takes two
shifts and an OR. Of the rotation amounts in the ChaCha20 double round (16, 12,
8, and 7 bits), the 16- and 8-bit rotations are byte-aligned, so we implement
them with \texttt{i8x16.shuffle}. The 12- and 7-bit rotations still use shifts
and ORs.

\subsection{AES-128 Kernel}

Our AES-128 kernel widens RustCrypto's 32-bit fixsliced
implementation~\cite{fixslicing} to 128-bit vectors so that eight registers
process eight blocks in parallel. Bitslicing is necessary because WASM-SIMD
has no AES instructions and cannot directly index a 256-entry S-box. The
32-bit implementation holds the state of two blocks in eight 32-bit words; we
widen each word to four lanes, so eight \texttt{v128} registers hold eight
blocks (\cref{fig:kernels}(b)). The S-box is computed with the Boyar--Peralta
circuit~\cite{boyar2012}, and the per-round changes of the state layout are
folded into the linear layer, so only a final layout correction remains instead
of repeated ShiftRows. Unlike auto-vectorized code, the kernel explicitly
implements the input/output transposition and the round operations; it shares
the existing key schedule. It covers AES-128 encryption, which suffices for CTR
mode in both directions; AES-256 is not implemented.

The kernel is constant-time at the algorithmic level: it uses no
secret-dependent branches or table lookups. It achieves this within standard
WASM-SIMD, without extensions such as Constant-time
WebAssembly~\cite{constanttime}. Whether the JIT-compiled code preserves this
property is a separate question (\cref{sec:discussion}).

\subsection{Scalar MAC Components}
\label{sec:mac}

The MACs, GHASH and Poly1305, remain scalar because our WASM-SIMD prototypes
of both were no faster than 64-bit scalar multiplication in preliminary
experiments. The 64-bit multiplications these MACs need do not map efficiently
to WASM-SIMD on our platform, and decomposing them into 32-bit multiplications
increases both the number of multiplications and the data rearrangement across
lanes. As a result, the MAC accounts for about 40\% of the cryptographic CPU
time of \cfg{Wasm-HandVec} with AES-128-GCM and about 50\% with
ChaCha20-Poly1305. WASM-SIMD does provide widening multiplications
(\texttt{i64x2.extmul\_low\_i32x4\_u} and its high counterpart), so multi-block
Poly1305 with narrow limbs and 32-by-32-bit products, as in
HACL$\times$N~\cite{polubelova2020}, may fare better than our prototypes; we
leave this to future work.

\section{Evaluation}
\label{sec:eval}

\begin{figure}[t]
  \centering
  \includegraphics[width=0.8\linewidth]{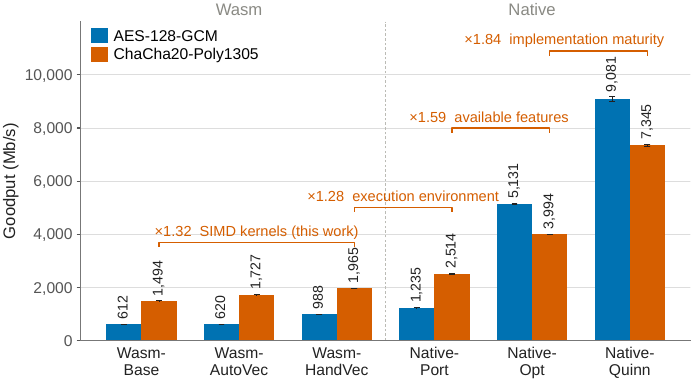}
  \caption{Receiver-side goodput for 1-GiB transfers over the 10GbE link. Bars
    show the mean of 20 runs per configuration, and error bars show 95\%
    confidence intervals based on the $t$ distribution. The sender is
    \cfg{Native-Quinn} for all configurations. The x-axis labels correspond to
    the configurations in \cref{fig:stack}. The ratios in the figure refer to
    ChaCha20-Poly1305.}
  \label{fig:main}
\end{figure}

\begin{figure}[t]
  \centering
  \includegraphics[width=0.95\linewidth]{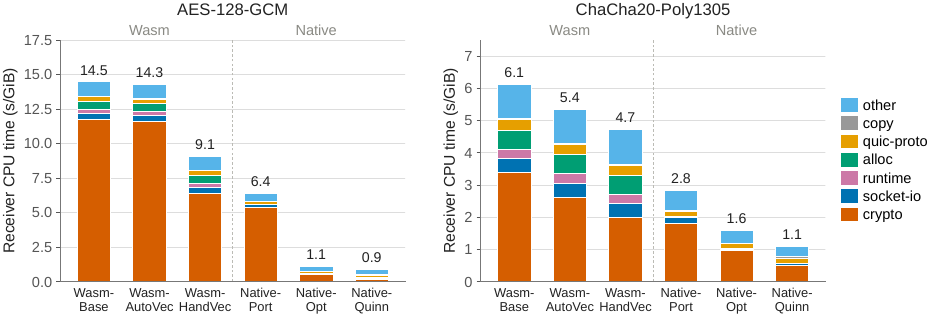}
  \caption{Receiver CPU time per GiB by category, drawn from \cref{tbl:cycles}.
    Numbers above the bars are totals in s/GiB. Each SIMD step shrinks the
    cryptographic part of the Wasm builds, whereas \texttt{alloc} and
    \texttt{runtime} nearly vanish only in the native builds.}
  \label{fig:cpu}
\end{figure}

Our kernels raise Wasm goodput substantially, yet a wide gap to native remains.
\Cref{fig:main} shows receiver goodput for each configuration, and
\cref{tbl:cycles,fig:cpu} break down receiver CPU time by category.

\begin{table}[t]
  \caption{Breakdown of receiver CPU time by category (ms/GiB). \emph{Total} is
    the mean CPU time over 20 runs; each category is the total multiplied by
    that category's share, averaged over 5 CPU profiles. \texttt{crypto}:
    cryptography; \texttt{copy}: memory copies; \texttt{socket-io}: socket
    processing; \texttt{runtime}: Wasmtime runtime processing; \texttt{alloc}:
    memory allocation and management; \texttt{quic-proto}: QUIC state updates
    and frame/header encoding and decoding; \texttt{other}: samples matching no
    category. The total also includes the omitted \texttt{ack-loss} category.
    The last column is crypto divided by total. Differences quoted in the text
    are computed from unrounded values.}
  \label{tbl:cycles}
  \centering
  \footnotesize
  \setlength{\tabcolsep}{3.2pt}
  \begin{tabular}{@{}llrrrrrrrrr@{}}
    \toprule
    Algorithm & Build & crypto & copy & socket-io & runtime & alloc & quic-proto & other & Total & crypto (\%) \\
    \midrule
     \multirow{6}{*}{\hd{AES-128-\\GCM}} & \cfg{Wasm-Base} & 11,768 & 38 & 441 & 281 & 593 & 322 & 1,038 & 14,483 & 81.3 \\
     & \cfg{Wasm-AutoVec} & 11,598 & 37 & 443 & 285 & 589 & 319 & 1,033 & 14,306 & 81.1 \\
     & \cfg{Wasm-HandVec} & 6,423 & 38 & 423 & 302 & 591 & 303 & 1,005 & 9,089 & 70.7 \\
     & \cfg{Native-Port} & 5,404 & 27 & 209 & 0.0 & 22 & 156 & 606 & 6,424 & 84.1 \\
     & \cfg{Native-Opt} & 534 & 14 & 36 & 0.5 & 24 & 141 & 389 & 1,138 & 46.9 \\
     & \cfg{Native-Quinn} & 184 & 37 & 89 & 0.0 & 22 & 170 & 399 & 902 & 20.4 \\
    \midrule
     \multirow{6}{*}{\hd{ChaCha20-\\Poly1305}} & \cfg{Wasm-Base} & 3,403 & 38 & 423 & 291 & 584 & 331 & 1,071 & 6,142 & 55.4 \\
     & \cfg{Wasm-AutoVec} & 2,631 & 33 & 439 & 284 & 594 & 315 & 1,057 & 5,355 & 49.1 \\
     & \cfg{Wasm-HandVec} & 2,004 & 34 & 425 & 287 & 585 & 319 & 1,073 & 4,726 & 42.4 \\
     & \cfg{Native-Port} & 1,804 & 29 & 203 & 0.0 & 22 & 161 & 618 & 2,837 & 63.6 \\
     & \cfg{Native-Opt} & 976 & 20 & 36 & 0.1 & 23 & 145 & 400 & 1,599 & 61.0 \\
     & \cfg{Native-Quinn} & 495 & 36 & 71 & 0.0 & 19 & 153 & 331 & 1,106 & 44.8 \\
    \bottomrule
  \end{tabular}
\end{table}

\subsection{Gains from WASM-SIMD}

WASM-SIMD cuts cryptographic CPU time while leaving the other categories
nearly unchanged (\cref{tbl:cycles}). From \cfg{Wasm-Base} to
\cfg{Wasm-HandVec}, goodput rises by 61.34\% with AES-128-GCM and by 31.55\%
with ChaCha20-Poly1305.

Auto-vectorization helps ChaCha20-Poly1305 but barely helps AES-128-GCM:
\cfg{Wasm-AutoVec} raises goodput over \cfg{Wasm-Base} by 15.60\% and 1.28\%,
respectively. ChaCha20 applies regular ARX operations to 32-bit words, which
the compiler maps to identical operations across lanes; auto-vectorization
alone cuts its cryptographic CPU time by 22.7\%. AES in \cfg{Wasm-Base}, in
contrast, already exploits bit-level parallelism within each word: the
RustCrypto \texttt{aes} crate uses a table-free fixsliced implementation for
32-bit platforms that spreads two blocks across eight 32-bit words and computes
the S-box with logic operations alone. Its linear layer consists of bit swaps
and rotations whose shift amounts differ from word to word, leaving the
compiler little to merge across lanes. GHASH, for its part, emulates carry-less
multiplication with constant-time 64-bit integer multiplication; WASM-SIMD has
no carry-less multiplication, and even its lane-wise \texttt{i64x2.mul}
expands into multiple instructions on x86-64.

Explicit kernels help both algorithms, and for AES-128-GCM they succeed where
auto-vectorization fails, showing that the state layout and operations must be
designed explicitly for WASM-SIMD. Relative to \cfg{Wasm-AutoVec}, the AES-128
kernel raises goodput by 59.31\% and cuts cryptographic CPU time by 44.6\%,
whereas the ChaCha20 kernel raises goodput by 13.79\% and cuts cryptographic
CPU time by 23.8\%. As Amdahl's law predicts, the AES-128-GCM gain is larger
because cryptography is both a larger share of its CPU time and reduced by a
larger fraction. For ChaCha20-Poly1305, auto-vectorization had already cut
cryptographic CPU time by 22.7\%, leaving less for the explicit kernel to
remove.

\subsection{Cipher Choice in Pure Wasm}

ChaCha20-Poly1305 remains the faster algorithm in pure Wasm even after our
kernels: \cfg{Wasm-HandVec} reaches 1,965.1~Mb/s with it but 988.1~Mb/s with
AES-128-GCM, a factor of 1.99. The factor shrinks from 2.44 at baseline because
our kernels accelerate AES more than ChaCha20 in relative terms. This result,
together with the 81.3\% cryptographic share of the AES baseline, is
consistent with Denis's finding that pure-Wasm AES-GCM is
slow~\cite{denis2026}. Without hardware AES, ChaCha20-Poly1305 is the better
choice for QUIC in our stack.

\subsection{Comparison Across Configurations}

The fastest native configuration still far outpaces the fastest Wasm one:
\cfg{Native-Quinn} delivers 9.19 times the goodput of \cfg{Wasm-HandVec} with
AES-128-GCM and 3.74 times with ChaCha20-Poly1305. With ChaCha20-Poly1305,
WASM-SIMD raises Wasm goodput from 20.3\% (\cfg{Wasm-Base}) to 26.8\%
(\cfg{Wasm-HandVec}) of \cfg{Native-Quinn} with the same cipher. Spies et
al.~\cite{spies2025} instead normalized by the fastest native configuration
regardless of cipher; by that measure, which here is \cfg{Native-Quinn} with
AES-128-GCM, \cfg{Wasm-Base} and \cfg{Wasm-HandVec} reach 16.4\% and 21.6\%.
Their measurements, however, were taken at the sender over a loopback path, so
the two results are not directly comparable.

We decompose the ChaCha20-Poly1305 gap into three successive goodput ratios,
each isolating one factor (\cref{tbl:decomp}): 1.28 from \cfg{Wasm-HandVec} to
\cfg{Native-Port}, 1.59 from \cfg{Native-Port} to \cfg{Native-Opt}, and 1.84
from \cfg{Native-Opt} to \cfg{Native-Quinn}.

\begin{table}[t]
  \caption{Decrease in receiver CPU time (ms/GiB) between successive
    configurations with ChaCha20-Poly1305, by category, computed from the
    unrounded values behind \cref{tbl:cycles}. Negative values are increases.
    Columns may not sum exactly because of rounding.}
  \label{tbl:decomp}
  \centering
  \small
  \begin{tabular}{@{}lrrr@{}}
    \toprule
     & \cfg{Wasm-HandVec} & \cfg{Native-Port} & \cfg{Native-Opt} \\
     & $\to$ \cfg{Native-Port} & $\to$ \cfg{Native-Opt} & $\to$ \cfg{Native-Quinn} \\
    \midrule
    \texttt{crypto} & 200 & 828 & 480 \\
    \texttt{copy} & 5 & 9 & $-15$ \\
    \texttt{socket-io} & 222 & 167 & $-35$ \\
    \texttt{runtime} & 287 & $-0.1$ & 0.1 \\
    \texttt{alloc} & 563 & $-0.7$ & 4 \\
    \texttt{quic-proto} & 158 & 17 & $-8$ \\
    \texttt{other} & 455 & 218 & 69 \\
    \texttt{ack-loss} & 1 & 0.0 & $-0.6$ \\
    \midrule
    Total decrease & 1,889 & 1,238 & 494 \\
    Goodput ratio & 1.28 & 1.59 & 1.84 \\
    CPU-time ratio & 1.67 & 1.77 & 1.45 \\
    \bottomrule
  \end{tabular}
\end{table}

The first ratio, 1.28, reflects the cost of the Wasm execution environment,
which lies mostly outside cryptography. Because \cfg{Wasm-HandVec} and
\cfg{Native-Port} share their code and SIMD width, their cryptographic CPU
times are close: 2,004 and 1,804~ms/GiB, about 10\% apart. \cfg{Wasm-HandVec}
instead spends more in every non-cryptographic category
(\cref{fig:cpu,tbl:decomp}). Of the 1,889~ms/GiB difference in total CPU time,
memory allocation accounts for 563~ms/GiB, \texttt{other} for 455, runtime
processing for 287, socket processing for 222, cryptography for 200, and QUIC
protocol processing for 158. Memory allocation and runtime processing, both
nearly absent in the native builds, explain nearly half of the difference.

The second ratio, 1.59, reflects host features that Wasm cannot use: wider SIMD
and kernel offloads. Of the 1,238~ms/GiB decrease, cryptography accounts for
828~ms/GiB, which with ChaCha20-Poly1305 can come only from AVX2, followed by
\texttt{other} (218) and socket processing (167), the expected effect of
GSO/GRO. Wasm cannot obtain the latter gain today: the UDP API of WASI~0.3
handles one datagram per send or receive call~\cite{wasisockets}, and
Wasmtime~49.0.2 likewise issues one system call per datagram. Narrowing this
gap requires WASI to expose host-side batching such as GSO/GRO, although, as
Jaeger et al.~\cite{jaeger2023} found natively, offloads also need matching
batched I/O in the application to pay off.

The third ratio, 1.84, stems almost entirely from the cryptographic libraries
of the two native implementations. \cfg{Native-Quinn} spends 495~ms/GiB on
cryptography, about half of the 976~ms/GiB of \cfg{Native-Opt}, and
cryptography accounts for 480 of the 494~ms/GiB decrease in total CPU time
(\cref{tbl:decomp}). The ring library of \cfg{Native-Quinn} and the RustCrypto
library of \cfg{Native-Opt} differ in their implementations and assembly
optimizations, which is the likely cause. Making a library available on a
platform and making it fast there are therefore separate problems.

\subsection{Estimating Further Gains with Amdahl's Law}
\label{sec:limit-theory}

Amdahl's law~\cite{amdahl1967} lets us estimate how much of the remaining gap
cryptography alone could close. Our kernels change mainly cryptographic CPU
time, unlike optimizations such as batched I/O that change \texttt{socket-io},
\texttt{alloc}, and \texttt{copy} at once, so we can model their effect while
holding other costs constant. Let $T$ be the receiver CPU time per GiB, $c$ the
share of cryptography in $T$, and $r$ the fraction by which cryptographic time
is reduced. The new CPU time is $T' = T - r\,c\,T$, so the predicted speedup is
\begin{equation}
  S = \frac{T}{T'} = \frac{1}{1 - c\,r}.
  \label{eq:amdahl}
\end{equation}
Because the Wasm receivers are CPU-bound (\cref{tbl:cpubound}), goodput is
inversely proportional to $T$, and the predicted goodput is $G' = S\,G$.

The model reproduces the measured gains from \cfg{Wasm-AutoVec} to
\cfg{Wasm-HandVec} within 2\%. It predicts speedups of 1.567 for AES-128-GCM
and 1.133 for ChaCha20-Poly1305, against measured 1.593 and 1.138.
Non-cryptographic CPU time stays nearly constant, changing from 2,708 to
2,665~ms/GiB with AES-128-GCM and from 2,724 to 2,722~ms/GiB with
ChaCha20-Poly1305. This agreement is a self-consistency check, since $c$ and
$r$ come from the same profiles, not an independent validation of the
category classification.

Even with the cryptographic CPU time of \cfg{Native-Opt}, which shares the
codebase but uses wider SIMD and AES instructions, Wasm goodput would reach
only about one third of \cfg{Native-Quinn}. We substitute \cfg{Native-Opt}'s
cryptographic CPU time into \cfg{Wasm-HandVec} and keep non-cryptographic time
constant. For ChaCha20-Poly1305, $c = 2004 / 4726 = 0.424$ and
$r = (2004 - 976) / 2004 = 0.513$, so \cref{eq:amdahl} predicts a speedup of
$1 / (1 - 0.424 \times 0.513) = 1.28$ and a goodput of 2,511~Mb/s: 62.9\% of
\cfg{Native-Opt} and 34.2\% of \cfg{Native-Quinn} with the same cipher. For
AES-128-GCM, the prediction is 2,807~Mb/s, or 54.7\% of \cfg{Native-Opt} and
30.9\% of \cfg{Native-Quinn}. In CPU time, the predicted 3,698~ms/GiB of
\cfg{Wasm-HandVec} with ChaCha20-Poly1305 is still 2.3 times the 1,599~ms/GiB
of \cfg{Native-Opt}; with AES-128-GCM the factor is 2.8.

These estimates are conditional, not upper bounds. Reducing non-cryptographic
costs could raise Wasm goodput further, whereas \cfg{Native-Opt}-level
cryptography is out of reach within WASM-SIMD, which offers neither AVX2-width
vectors nor AES instructions. Because the native receivers are not saturated
(\cref{tbl:cpubound}), comparisons with their measured goodput also understate
the remaining gap.

The non-cryptographic costs that remain point to memory allocation, runtime
processing, and the use of host features as the next optimization targets. The
large \texttt{other} category needs finer analysis before its cost can be
attributed to specific causes.

\section{Discussion and Limitations}
\label{sec:discussion}

\paragraph{Strength of the baseline.}
Our AES gain is measured against RustCrypto's default \texttt{wasm32} path,
which may not be the strongest scalar baseline. That path is a fixsliced AES
with 32-bit words and two blocks per call. On 64-bit targets the same crate
selects a variant with 64-bit words and four blocks, which could be faster in
Wasm because Wasm has native 64-bit integer operations; we have not evaluated
it. The 61.34\% gain for AES-128-GCM could therefore be smaller against a
64-bit scalar variant. Likewise, \cfg{Wasm-AutoVec} reflects what the compiler
achieves on this code, not what any auto-vectorizer could achieve.

\paragraph{Scope of the cipher comparison.}
Our cipher comparison applies to our RustCrypto-based stack, not to AES-GCM in
Wasm in general. In our stack, GHASH is scalar and the AES kernel covers only
AES-128. The results do not show that AES-GCM is intrinsically slower than
ChaCha20-Poly1305 in Wasm and say nothing about AES-256-GCM; a faster GHASH
could narrow the gap.

\paragraph{Native configurations are not CPU-bound.}
The goodput ratios between Wasm and native understate the efficiency gap
because the native receivers are not CPU-bound. \cfg{Native-Port} and
\cfg{Native-Opt} use only 68--92\% of the transfer time, and with
\cfg{Native-Quinn} and ChaCha20-Poly1305 the sender reaches up to 88.8\% CPU
utilization and may partly limit goodput. Accordingly,
the first two CPU-time ratios in \cref{tbl:decomp} (1.67 and 1.77) exceed the
corresponding goodput ratios (1.28 and 1.59). We report goodput ratios in
\cref{sec:eval} because goodput is what an application observes.

\paragraph{Fixed per-process costs.}
Costs incurred once per process inflate our 1-GiB Wasm figures. The nearly
constant 0.35--0.46~s excess of CPU time over transfer time in the Wasm builds
suggests a fixed cost of roughly 0.4~s, such as module compilation, similar in
magnitude to the \texttt{runtime} category (281--302~ms/GiB). If so, the
\texttt{runtime} category and part of \texttt{other} would amortize over longer
transfers, and the steady-state gap to native would be somewhat smaller. We
have not verified this with longer transfers.

\paragraph{Statistical reporting.}
The category breakdown is less certain than the goodput measurements. Goodput
is averaged over 20 runs with 95\% confidence intervals (\cref{fig:main}),
whereas the breakdown averages 5 profiles without confidence intervals, so
differences of a few percent in a single category should not be
over-interpreted. The name-based classification may also misclassify samples.
In particular, \texttt{other} includes unmatched OS-kernel processing and
accounts for 22.7\% of total CPU time in \cfg{Wasm-HandVec} with
ChaCha20-Poly1305; breaking it down is necessary to identify which costs are
specific to Wasm.

\paragraph{Platform and workload.}
Our results cover one hardware platform, one runtime, and one workload: 1-GiB
bulk transfers between high-performance x86-64 machines running Wasmtime
49.0.2 over a direct, lossless, low-latency link. The benefit of SIMD depends
on the number of vector execution units, the clock frequency, and the cache
hierarchy, so validation is needed on Internet of Things (IoT) and edge
devices, on other instruction sets and runtimes, and over paths that emulate
real networks. For short connections, the handshake and header protection
account for a larger share of processing, so the gains may be smaller. The
kernels also process packet-sized inputs of up to roughly 1,400 bytes, for
which the handling of the last few blocks of each packet
matters~\cite{oku2020}. Given the performance difference between the
experimental rustls-rustcrypto and ring, evaluation with multiple QUIC
implementations is also needed.

\paragraph{Runtime dependence.}
The absolute gains depend on how the runtime lowers WASM-SIMD. Our
byte-shuffle rotations and 128-bit kernels rely on Wasmtime lowering
\texttt{i8x16.shuffle} and other vector instructions to efficient x86-64
instructions. Other runtimes, such as browser engines, and other architectures
may lower them differently, although the design principles carry over.

\paragraph{Single guest thread and packet size.}
Our setup measures per-core efficiency at a 1,500-byte MTU and likely
understates the importance of the kernels at larger MTUs. We run one
connection on one guest thread; scaling across cores, for example with one
connection per core, is orthogonal to this efficiency. A larger MTU would amortize per-packet
non-cryptographic costs~\cite{koenig2025} and likely raise the share of
cryptography, making the kernels more important; we did not test this.

\paragraph{Host-provided cryptography.}
Our results set the bar that host-provided cryptography such as WASI-Crypto
would have to beat, but we did not measure it. Such an interface would exploit
AES instructions and wider vectors but incur guest--host transitions and memory
copies for every packet; a direct comparison is future work.

\paragraph{Side channels.}
Whether the constant-time design survives JIT compilation remains unverified.
The kernels avoid secret-dependent branches and table lookups, but future work
must evaluate resistance to timing attacks and explore integration with
Constant-time WebAssembly~\cite{constanttime}.

\paragraph{Practical relevance of the absolute numbers.}
Our best Wasm configuration, at about 2~Gb/s, remains far below the 10GbE line
rate and reaches 27\% of \cfg{Native-Quinn} with the same cipher. This paper
contributes a quantified path toward closing the gap, not a claim that it has
been closed.

\section{Conclusion and Future Work}
\label{sec:conclusion}

Explicit WASM-SIMD kernels substantially accelerate a pure-Wasm QUIC stack,
but cryptographic acceleration alone cannot close the gap to native. Our
eight-block fixsliced AES and four-block ChaCha20 kernels raise goodput in a
WASI~0.3 QUIC stack by 61.34\% and 31.55\% over the non-SIMD baseline, and
the comparison with auto-vectorization shows that state layouts and operations
must be designed explicitly for the constraints of WASM-SIMD. After
optimization, ChaCha20-Poly1305 delivers 1.99 times the goodput of
AES-128-GCM, confirming end to end the importance of cipher choice suggested by
stand-alone benchmarks~\cite{denis2026}. Profiling and Amdahl's law predict
that, even with \cfg{Native-Opt}-level cryptography and unchanged
non-cryptographic costs, goodput would reach only 34.2\% (ChaCha20-Poly1305)
and 30.9\% (AES-128-GCM) of \cfg{Native-Quinn} with the same cipher. The next
targets are therefore memory allocation, runtime overhead, and efficient use of
host features such as batched sends and receives~\cite{aoyagi2026}.

The scalar components and untested baselines define the immediate future work.
GHASH, Poly1305, and the single-block AES-ECB encryption for header protection
remain scalar; the MACs alone account for 40--50\% of the cryptographic CPU
time of \cfg{Wasm-HandVec}, and our SIMD prototypes did not reduce their cost,
so their arithmetic structure must be reconsidered. Experiments that would
strengthen our conclusions include a 64-bit fixsliced scalar baseline, longer
transfers, a comparison with WASI-Crypto~\cite{wasicrypto} that accounts for
guest--host transitions and copies, and an ablation of the individual
\cfg{Native-Opt} features (AVX2, GSO/GRO, and AES-NI).

\section*{Acknowledgments}
This work was supported by the FORWARD commissioned research program of the
Ministry of Internal Affairs and Communications (MIC), Japan
(JPMI260720003).


\end{document}